\documentclass[floatfix,twocolumn,showpacs,amssymb,amsmath,prd]{revtex4}
\usepackage{graphicx}
\usepackage{dcolumn}
\usepackage{bm}

\begin{document}
\title{The primordial ``$f_{\mathrm{NL}}$'' non-Gaussianity, and perturbations beyond the present horizon}
\author{Jaiseung Kim}
\email{jkim@nbi.dk}
\affiliation{Niels Bohr Institute, Blegdamsvej 17, DK-2100 Copenhagen, Denmark}
\author{Pavel Naselsky}
\affiliation{Niels Bohr Institute, Blegdamsvej 17, DK-2100 Copenhagen, Denmark}

\date{\today}

\begin{abstract}
We show a primordial non-linear (``$f_{\mathrm{NL}}$'') term may produce unphysically large CMB anisotropy for a red-tilted primordial power spectrum ($n<1$), because of coupling to primordial fluctuation on the largest scale. 
We consider a primordial power spectrum models of a running spectral index, and a transition at very low wavenumbers.
We find that  only negative running spectral index models are allowed, provided that there is no transition at a low wavenumbers (i.e. $k\ll 1$).
For models of a constant spectral index, we find $\log(k_c/k_0)\gtrsim -184$, at $1\sigma$ level, on the transition scale of sharp cut-off models, using recent CMB and SDSS data.
\end{abstract}


\maketitle 

\section{Introduction}

Recently, there have been great successes in measurement of Cosmic Microwave Background (CMB) anisotropy by ground and satellite observations \citep{WMAP5:basic_result,WMAP5:powerspectra,WMAP5:parameter,ACBAR,ACBAR2008,QUaD1,QUaD2,QUaD:instrument}.
The five year data of the Wilkinson Microwave Anisotropy Probe (WMAP) \cite{WMAP5:basic_result,WMAP5:powerspectra,WMAP5:parameter} is released and the recent ground-based CMB observations such as the ACBAR \cite{ACBAR,ACBAR2008} and QUaD \cite{QUaD1,QUaD2,QUaD:instrument} provide information complementary to the WMAP data. In near future, PLANCK surveyor \cite{Planck:sensitivity,Planck:mission} is going to measure CMB temperature and polarization anisotropy with great accuracy over wide range of angular scales. Using the observational data, we are able to impose strong constraints on cosmological models \citep{Modern_Cosmology,Inflation,Foundations_Cosmology}, and in particular,
on the class of inflation models of very large number of e-folds $N_e\gg 100$. They may provide a new window on physics beyond the Planck scale \cite{Inflation_Planckian_problem, Inflation_Planckian_slowroll}.
Another important feature of the recent CMB observations is testing non-Gaussianity and statistical anisotropy of the CMB sky \citep{cold_spot1,cold_spot2,cold_spot_wmap3,cold_spot_origin,Tegmark:Alignment,Multipole_Vector,Multipole_Vector2,Axis_Evil,Axis_Evil2,Axis_Evil3,Power_Asymmetry,power_asymmetry_subdegree,power_asymmetry_wmap5}. They provide a unique opportunity to test the modern theories of inflation through the
observational data (see for review \cite{Non_Gaussianity}). 

The fluctuations of the gravitational potential $\Phi(\mathbf{x})$ (equivalent to  Bardeen's gauge invariant variable $\Phi_{H}$ \citep{Bardeen})
is related to  primordial perturbation in complicated ways \citep{Maldacena_fnl,WMAP5:Cosmology}.
When considered up to the second order, there exists a nonlinear term $f_{\mathrm{NL}}\,\Phi^2_{\mathrm{L}}(\mathbf{x})$, where $f_{\mathrm{NL}}$ is a coupling constant. The nonlinear term $f_{\mathrm{NL}}\,\Phi^2_{\mathrm{L}}(\mathbf{x})$ leads to coupling between largest scales and scales relevant to observable Universe. 
The recent constraint of the WMAP data shows $f_{\mathrm{NL}}\sim 60\pm 30$ (see \cite{WMAP3:Gaussianity,fnl_Yadav,fnl_Smith,WMAP5:Cosmology,fnl_Smith} for the recent analysis). 

Much of studies have been focused on the behavior of the primordial power spectrum on small scales. In this paper, we focus on the `infrared' asymptotic behavior of a red-tilted ($n<1$) primordial power spectrum.
Given the red-tilted primordial power spectra \cite{WMAP3:parameter,WMAP5:Cosmology}, coupling to the fluctuation on largest scales may produce unphysically large CMB anisotropy, which could be in disagreement with CMB observational data.
There have been attempts to remove the divergence of the nonlinear term by renormalization \citep{fnl_bias}, and the author notes that
the residual k-dependent term, which is not removed by renormalization, is negligible on observable scales for galaxy surveys.
Unlike galaxy surveys, the residual k-dependent term produce very large excess power on CMB anisotropy of low multipoles.
%, since CMB horizon is much larger than the scales of galaxy surveys.
%as also noted by the author, 
%this singularity is not removed completly by renormalization and k-dependent term is not removed by t
Not to produce unphysical excess power for CMB anisotropy, we require a primordial power spectrum to satisfy: 1) a spectral index of negative running or 2) a transition at very large scale (e.g. sharp cutoff in the power spectrum at a very low wavenumber). We find  at least one of them should be satisfied to make agreement with the recent CMB observational data.
As will be discussed in this paper, the imprints of the largest scales due to $f_{\mathrm{NL}}\,\Phi^2_{\mathrm{L}}(\mathbf{x})$ term may improve our understanding on the properties of primordial perturbations on the scales larger than the present particle horizon.    

The outline of this paper is as follows.
In Section \ref{nonlinear_term}, we discuss the primordial power spectrum associated with a primordial nonlinear (`$f_{\mathrm{NL}}$') term.
In Section \ref{CMB}, we discuss the effect of a `$f_{\mathrm{NL}}$' term on CMB power spectra. In Section \ref{Primordial_Power}, we show the primordial power spectrum should satisfy some requirement not to produce unphysically large CMB power spectra.
In Section \ref{Discussion}, we summarize our investigation and discuss prospects. 

\section{the effect of the ``$f_{\mathrm{NL}}$'' term on a primordial power spectrum}
\label{nonlinear_term}
Up to second order, primordial perturbation is given by:
\citep{CMB_fnl,Komatsu_thesis,fnl_simulation,WMAP5:Cosmology}:
\begin{eqnarray}
\Phi(\mathbf x)=\Phi_{\mathrm L}(\mathbf x) + f_{\mathrm {NL}}\left[\Phi^2_{\mathrm L}(\mathbf x)-\langle \Phi^2_{\mathrm L}(\mathbf x) \rangle \right],\label{Phi_real}
\end{eqnarray}
where $\Phi_{\mathrm L}(\mathbf x)$ is a linear part of primordial perturbation, and $f_{\mathrm {NL}}$ is the non-linear coupling parameter. 
The last term on the right hand side ensures $\langle\Phi(\mathbf x)\rangle=0$, and is given by:
\begin{eqnarray*}
\langle \Phi^2_{\mathrm L}(\mathbf x) \rangle &=&\int P_{\Phi}(k)\,\frac{d^3\mathbf k}{(2\pi)^3},
\end{eqnarray*}
where
\begin{eqnarray*}
P_{\Phi}(k)=\frac{\Delta^2_{\mathcal R}(k)}{k^3}.
\end{eqnarray*}
$\Delta^2_{\mathcal R}(k)$ is the variance of curvature perturbation per logarithmic interval 
$d\ln k$ \citep{Inflation,WMAP5:Cosmology}.
Using Eq. \ref{Phi_real}, we find primordial perturbation in Fourier space:
\begin{eqnarray}
\Phi(\mathbf k)=\Phi_{\mathrm L}(\mathbf k)+\Phi_{\mathrm {NL}}(\mathbf k),\label{Phi_Fourier}
\end{eqnarray}
where
\begin{eqnarray}
\lefteqn{\Phi_{\mathrm {NL}}(\mathbf k)=}\label{Phi_NL}\\
&&f_{\mathrm {NL}}\left(\int \Phi_{\mathrm L}(\mathbf k+\mathbf p) \Phi^*_{\mathrm L}(\mathbf p)
\,\frac{d^3\mathbf p}{(2\pi)^3} -(2\pi)^3\delta(\mathbf k)\langle \Phi^2_{\mathrm L}(\mathbf x) \rangle \right).\nonumber
\end{eqnarray}
In most of inflationary models, 
$\Phi_{\mathrm L}(\mathbf k)$ follows a Gaussian distribution \citep{Modern_Cosmology,Inflation,Foundations_Cosmology,Komatsu_thesis,fnl_simulation,WMAP5:Cosmology}, and hence have the following statistical properties:
\begin{eqnarray}
\langle \Phi_{\mathrm L}(\mathbf k)\rangle &=&0,\label{Phi_mean}\\
\langle \Phi_{\mathrm L}(\mathbf k) \Phi^*_{\mathrm L}(\mathbf k') \rangle &=&(2\pi)^3 P_{\Phi}(k)\,\delta(\mathbf k-\mathbf k'),\label{Phi_correlation}\\
\langle \Phi_{\mathrm L}(\mathbf k+\mathbf p) \Phi^*_{\mathrm L}(\mathbf p) \rangle&=&(2\pi)^3 \delta(\mathbf k) \,P_{\Phi}(p),\label{Phi_twocorrelation}
\end{eqnarray}
\begin{eqnarray}
\lefteqn{\langle \Phi_{\mathrm L}(\mathbf k+\mathbf p) \Phi^*_{\mathrm L}(\mathbf p) 
\,\Phi^*_{\mathrm L}(\mathbf k'+\mathbf p') \Phi_{\mathrm L}(\mathbf p')\rangle=(2\pi)^6\times}\nonumber\\
&&\left[P_{\Phi}(k+p)P_{\Phi}(p)\,\delta(\mathbf k-\mathbf k')\left(\delta(\mathbf p-\mathbf p')+\delta(\mathbf k+\mathbf p+\mathbf p')\right)\right.\nonumber\\
&&\left.+\delta(\mathbf k)\,\delta(\mathbf k')\,P_{\Phi}(p)\,P_{\Phi}(p')\right]\label{Phi_fourcorrelation}
\end{eqnarray}
Using Eq. \ref{Phi_NL}, \ref{Phi_mean}, \ref{Phi_correlation}, \ref{Phi_twocorrelation} and \ref{Phi_fourcorrelation}, 
we may easily show
\begin{eqnarray}
\langle \Phi_{\mathrm L}(\mathbf k) \Phi^*_{\mathrm {NL}}(\mathbf k') \rangle&=&0,\label{LNL}\\
\langle \Phi_{\mathrm {NL}}(\mathbf k) \Phi^*_{\mathrm {NL}}(\mathbf k') \rangle&=&8\pi\,(f_{\mathrm{NL}})^2\,\delta(\mathbf k-\mathbf k')\nonumber\\ 
&\times& \left[\int P_\Phi(k+p)\,P_\Phi(p)\,{p}^2 dp\right].\label{NLNL}
\end{eqnarray}
Finally, by using Eq. \ref{Phi_Fourier}, \ref{Phi_correlation}, \ref{LNL} and \ref{NLNL} we find: 
\begin{eqnarray*}
\langle \Phi^*(\mathbf k) \Phi(\mathbf k') \rangle =(2\pi)^3 \left[P_{\Phi}(k)+P_{\Phi,\mathrm{NL}}(k)\right]
\delta(\mathbf k-\mathbf k'),
\end{eqnarray*}
where
\begin{eqnarray}
P_{\Phi,\mathrm{NL}}(k)&=&\frac{(f_{\mathrm{NL}})^2}{\pi^2}\int P_\Phi(k+k') P_\Phi(k')\:{k'}^2 dk'.\label{P_conv}
\end{eqnarray}

\section{the effect of the `$f_{\mathrm{NL}}$' term on CMB power spectra}
\label{CMB}
The Stokes parameters of CMB anisotropy are conveniently decomposed in terms of spin $0$ and spin $\pm2$ spherical harmonics: \begin{eqnarray*}
T(\hat{\mathbf n})&=&\sum_{lm} a_{T,lm}\,Y_{lm}(\hat{\mathbf n}),\\
Q(\hat {\mathbf n})\pm i U(\hat {\mathbf n})&=&\sum_{l,m} -(a_{E,lm}\pm i \,a_{B,lm})\;{}_{\pm2}Y_{lm}(\hat {\mathbf n}), 
\end{eqnarray*}
where $a_{T,lm}$, $a_{E,lm}$ and $a_{B,lm}$ are decomposition coefficients. 
The decomposition coefficients are related to primordial perturbations as:
\begin{eqnarray}
a_{T,lm}&=&4\pi (-\imath)^l \int \frac{d^3\mathbf k}{(2\pi)^3} \Phi(\mathbf k)\,g_{Tl}(k)\,Y^*_{lm}(\hat {\mathbf k}),\\
a_{E,lm}&=&4\pi (-\imath)^l \int \frac{d^3\mathbf k}{(2\pi)^3} \Phi(\mathbf k)\,g_{El}(k)\,Y^*_{lm}(\hat {\mathbf k}),\\
a_{B,lm}&=&4\pi (-\imath)^l \int \frac{d^3\mathbf k}{(2\pi)^3} \Phi(\mathbf k)\,g_{Bl}(k)\,Y^*_{lm}(\hat {\mathbf k}),
\end{eqnarray}
where $g_{Tl}(k)$, $g_{El}(k)$ and $g_{Bl}(k)$ are the radiation transfer functions and
can be numerically computed by a computer software \texttt{CAMB} \citep{CAMB}.
In the absence of tensor perturbation, CMB power spectra are given by:
\begin{eqnarray}
C^{TT}_{l}&=&\frac{2}{\pi} \int k^2 dk \left[P_{\Phi}(k)+P_{\Phi,\mathrm{NL}}(k)\right]g^2_{Tl}(k), \label{ClTT}\\
C^{EE}_{l}&=&\frac{2}{\pi} \int k^2 dk \left[P_{\Phi}(k)+P_{\Phi,\mathrm{NL}}(k)\right]g^2_{El}(k),\label{ClEE}\\
C^{TE}_{l}&=&\frac{2}{\pi} \int k^2 dk \left[P_{\Phi}(k)+P_{\Phi,\mathrm{NL}}(k)\right]g_{Tl}(k)g_{El}(k),\nonumber\\\label{ClTE}
\end{eqnarray}
where $P_{\Phi,\mathrm{NL}}(k)$ is the primordial power spectrum associated with the `$f_{\mathrm{NL}}$' term and given by Eq. \ref{P_conv}. Note that CMB anisotropy, excluding the dipole, is sensitive to primordial perturbation of wavenumbers $k\gtrsim 2/\eta_0$, where $\eta_0$ is the present conformal time.

\section{The shape of a Primordial Power Spectrum}
\label{Primordial_Power}
Inflation models predict the power spectrum of primordial perturbation nearly follow a power law \citep{CMB_nr,Modern_Cosmology,Inflation,Foundations_Cosmology,WMAP1_Inflation,WMAP3:parameter,WMAP5:Cosmology,WMAP5:parameter}.
Since fluctuations, which were once on sub-Planckian scales, are stretched to the observable scales by inflation, we need to consider trans-Planckian effects on a primordial power spectrum \citep{Inflation_Planckian_problem,Inflation_Planckian_spectra,Inflation_Planckian_note,Inflation_Planckian_estimate,CMB_Planckian_signature,WMAP_oscillation,Inflation_Planckian,Inflation_initial}.
Since trans-Planckian corrections are highly model-dependent \citep{Planckian_Astrophysics,CMB_Planckian_observation}, we consider general forms of trans-Planckian correction. Following the approach of the WMAP team, we model the variance of curvature perturbation $\Delta^2_{\mathcal R}(k)$ by two general forms \citep{WMAP3:parameter}:
\begin{eqnarray}
\Delta^2_{\mathcal R}(k)=A_0\left(\frac{k}{k_0}\right)^{n-1}\left[1+\epsilon_{\mathrm{TP}}\cos\left(v\frac{k}{k_0}+\phi\right)\right]\label{R1}
\end{eqnarray}
\begin{eqnarray}
\Delta^2_{\mathcal R}(k)=A_0
\left(\frac{k}{k_0}\right)^{n-1}\left[1+\epsilon_{\mathrm{TP}}\cos\left(v\ln{\frac{k}{k_0}}+\phi\right)\right]\label{R2}
\end{eqnarray}
where the pivot scale $k_0$ is set to the WMAP team's pivot scale $0.002/\mathrm{Mpc}$ \citep{WMAP3:parameter},
and the spectral index $n$ is given by:
\begin{eqnarray*}
n=n(k_0)+\frac{1}{2}\frac{d\,n}{d\ln k}\,\ln\left(\frac{k}{k_0}\right).
\end{eqnarray*}
$\epsilon_{TP}$, $v$, and $\phi$ are the amplitude, the frequency and the phase of trans-Planckian effect.
We denote the spectrum in Eq. \ref{R1} and \ref{R2} respectively as `the model I' and `the model II', which differ in the parametrized form of  trans-Planckian corrections.
Using Eq. \ref{P_conv}, \ref{R1} and \ref{R2}, we find 
\begin{eqnarray}
\lefteqn{P_{\Phi,\mathrm{NL}}(k)=\frac{(f_{\mathrm{NL}})^2}{\pi^2} A^2_0}\nonumber\\
&\times&\int^{\infty}_{0}\frac{dk'}{k_0}\left(\frac{k+k'}{k_0}\right)^{n-4}[1+\epsilon_{\mathrm{TP}}\cos\theta(k+k')]\nonumber\\ 
&\times&\left(\frac{k'}{k_0}\right)^{n-2} [1+\epsilon_{\mathrm{TP}}\cos\theta(k')]\label{P_NL}
\end{eqnarray}
where $\theta(k)=v\frac{k}{k_0}+\phi$ for the model I and $\theta=v\ln{\frac{k}{k_0}}+\phi$ for the model II.

Most of inflationary models predict that a primordial spectrum is slightly red-tilted (i.e. $n(k_0)<1$) \citep{Inflation,Foundations_Cosmology}, which is in good agreement with observations \citep{WMAP5:Cosmology}.
%Given a slightly red-tiltled spectral index $n<1$, $P_{\Phi,\mathrm{NL}}(k)$ shown in Eq. \ref{P_NL} may become quite large for $k\ll 1$.
%Since it has k-dependency, large excess power due to $P_{\Phi,\mathrm{NL}}(k)$ cannot be removed by ren
%Given a slightly red-tiltled spectral index $n<1$, $P_{\Phi,\mathrm{NL}}(k)$ shown in Eq. \ref{P_NL} may become large, therefore producing unphysically large CMB 
Given a slightly red-tiltled spectral index $n<1$, $P_{\Phi,\mathrm{NL}}(k)$ shown in Eq. \ref{P_NL} may become quite large. It increases with decreasing $k$ and 
has strong $k$-dependenc. This $k$-dependent excess power is not simply removed by renormalization, and may produce very large CMB power spectra on low multipoles (refer to Eq. \ref{ClTT}, \ref{ClEE} and \ref{ClTE}). 
%Given a slightly red-tiltled spectral index $n<1$, $P_{\Phi,\mathrm{NL}}(k)$ shown in Eq. \ref{P_NL} may become quite large, therefore producing very large CMB power %spectra (see Eq. \ref{ClTT}, \ref{ClEE} and \ref{ClTE}). 
CMB power spectra are well measured by recent satelite and ground observations \citep{ACBAR,QUaD_review,QUaD1,QUaD2,QUaD:instrument,WMAP5:basic_result,WMAP5:powerspectra}.
Not to produce unphysical large excess power, a primordial power spectrum should satisfy some condition, which will be discussed in the following subsections.

\subsection{running spectral index}
We consider a running spectral index (i.e. ${d\,n}/{d\ln k}\ne0$).
Since significant contribution to the integral comes from $k'/k_0\ll 1$,  we find $P_{\Phi,\mathrm{NL}}(k)$ for $k\gtrsim 2/\eta_0$ and the model I:
\begin{eqnarray}
\lefteqn{P_{\Phi,\mathrm{NL}}(k)\approx\frac{(f_{\mathrm{NL}})^2}{\pi^2} A^2_0 \left(\frac{k}{k_0}\right)^{n-4}\left[1+\epsilon_{\mathrm{TP}}\cos\left(v\frac{k}{k_0}+\phi\right)\right]}\nonumber\\
&\times&\int^{\infty}_{0} 
\left(1+\epsilon_{\mathrm{TP}}\cos\phi-\epsilon_{\mathrm{TP}}\,v\,\sin\phi\,\frac{k'}{k_0}\right)\left(\frac{k'}{k_0}\right)^{n-2} 
\frac{dk'}{k_0}.\nonumber\\
\label{P_NL1}
\end{eqnarray}
Note that we have set $k_{\mathrm{max}}$ to $\infty$, because the integrand converges to zero for $k'/k_0 \gg 1$.
If ${d\,n}/{d\ln k}<0$,
Eq. \ref{P_NL1} is given by
\begin{eqnarray*}
P_{\Phi,\mathrm{NL}}(k)&\approx&\frac{(f_{\mathrm{NL}})^2}{\pi^{2}} A^2_0 \left(\frac{k}{k_0}\right)^{n-4}\left[1+\epsilon_{\mathrm{TP}}\cos\left(v\frac{k}{k_0}+\phi\right)\right]\\
&\times& \sqrt{\frac{2\pi}{-\alpha}}\left[\left(1+\epsilon_{\mathrm{TP}}\cos\phi\right)\exp\left(-\frac{(n-1)^2}{2\alpha}\right)\right.\\
&&\left.-\epsilon_{\mathrm{TP}}\,v\,\sin\phi \exp\left(-\frac{n^2}{2\alpha}\right)\right], 
\end{eqnarray*}
where $\alpha={d\,n}/{d\ln k}$.
On the other hand, if ${d\,n}/{d\ln k}\ge 0$ and the lower integration bound $k_{\mathrm{min}}\to 0$,
Eq. \ref{P_NL1} approach an infinity: $P_{\Phi,\mathrm{NL}}(k) \to \infty$.
Hence, we see that ${d\,n}/{d\ln k}<0$ is required to keep CMB power spectra finite.

For $k\gtrsim 2/\eta_0$ and the model II, we find:
\begin{eqnarray*}
\lefteqn{P_{\Phi,\mathrm{NL}}(k)\approx\frac{(f_{\mathrm{NL}})^2}{\pi^2} A^2_0 \left(\frac{k}{k_0}\right)^{n-4}}\\
&\times&\left[1+\epsilon_{\mathrm{TP}}\cos\left(v\log\frac{k}{k_0}+\phi\right)\right]\\
&\times&\int^{\infty}_{0} 
\left[1+\epsilon_{\mathrm{TP}}\cos\left(v\log\frac{k}{k_0}+\phi\right)\right]\left(\frac{k'}{k_0}\right)^{n-2} 
\frac{dk'}{k_0}\\
&=&\frac{(f_{\mathrm{NL}})^2}{\pi^2} A^2_0 \left(\frac{k}{k_0}\right)^{n-4}\left[1+\epsilon_{\mathrm{TP}}\cos\left(v\log\frac{k}{k_0}+\phi\right)\right]\\
&\times&\int^{\infty}_{-\infty} 
\left[1+\epsilon_{\mathrm{TP}}\cos\left(vx +\phi\right)\right]e^{x(n-1)}dx,\\
\end{eqnarray*}
where $x=\log(k'/k_0)$.
We have also set $k_{\mathrm{max}}$ to $\infty$, because the integrand converges to zero for $k'/k_0 \gg 1$.

If ${d\,n}/{d\ln k}<0$, we get 
\begin{eqnarray*}
P_{\Phi,\mathrm{NL}}(k)&\approx&\frac{(f_{\mathrm{NL}})^2}{\pi^{2}} A^2_0 \left(\frac{k}{k_0}\right)^{n-4}\left[1+\epsilon_{\mathrm{TP}}\cos\left(v\frac{k}{k_0}+\phi\right)\right]\\
&\times& \sqrt{\frac{2\pi}{-\alpha}}
\exp\left(-\frac{(n-1)^2}{2\alpha}\right)\\
&\times&\left[1+\epsilon_{\mathrm{TP}}\cos\left(\phi-\frac{(n-1)v}{\alpha}\right) \exp\left(\frac{v^2}{2\alpha}\right)\right],
\end{eqnarray*}
where $\alpha={d\,n}/{d\ln k}$.
On the other hand, if ${d\,n}/{d\ln k}\ge 0$ and the lower integration bound $k_{\mathrm{min}}\to 0$,
Eq. \ref{P_NL1} also approach an infinity: $P_{\Phi,\mathrm{NL}}\to \infty$.
Therefore, we require ${d\,n}/{d\ln k}< 0$ in running spectral index models.
Many inflationary models predict a negative ${d\,n}/{d\ln k}$, and hence meet the requirement, while
a few inflationary models fail to satisfy the requirement. For instance, the model of a mass term potential $V(\phi)/V_0=1\pm 4\pi G \,|c|\,\phi^2$ predicts ${d\,n}/{d\ln k}=0$, and the model of the potential $V(\phi)/V_0=1-4\pi G |c| \phi^2\,\ln \phi/|Q|$, which belongs to the class of a softly broken SUSY models, predicts ${d\,n}/{d\ln k}>0$.
Therefore, these models are in disagreement with observations, provided the primordial power spectrum in Eq. \ref{R1} and
\ref{R2} are valid up to the largest spatial scale.

\subsection{sharp cut-off in the primordial power spectrum}
We consider a model of a constant spectral index (i.e. ${d\,n}/{d\ln k}=0$), and discuss some requirement to avoid unphysically large $P_{\mathrm{NL}}$.
We may consider a transition in the shape of the primordial power spectrum at very low wavenumber, below which Eq. \ref{R1} or \ref{R2} are no longer valid. For instance, the WMAP team have considered a model of a sharp cutoff, and found that the cut-off at $k_c\sim 3\times 10^{-4}/\mathrm{Mpc}$ makes a slightly better fit \citep{WMAP3:parameter}.
For $k>k_c$, we may write Eq. \ref{P_NL} as follows:
\begin{eqnarray}
\lefteqn{P_{\Phi,\mathrm{NL}}(k)=\frac{(f_{\mathrm{NL}})^2 A_0}{\pi^2}\left(\int^{k_c}_{0}\frac{dk'}{k_0}\left(\frac{k+k'}{k_0}\right)^{n-4}\right.}\nonumber\\
&\times&[1+\epsilon_{\mathrm{TP}}\cos\theta(k+k')]\;P_{\Phi}(k')\left(\frac{k'}{k_0}\right)^2\nonumber\\ &+&A_0\int^{k_{max}}_{k_c}\frac{dk'}{k_0}\left(\frac{k+k'}{k_0}\right)^{n-4}[1+\epsilon_{\mathrm{TP}}\cos\theta(k+k')]\nonumber\\ &\times&\left.\left(\frac{k'}{k_0}\right)^{n-2} [1+\epsilon_{\mathrm{TP}}\cos\theta(k')]\right)\label{P_NL2}
\end{eqnarray}
where $\theta(k)=v\frac{k}{k_0}+\phi$ for the model I and $\theta=v\ln{\frac{k}{k_0}}+\phi$ for the model II.

Just as the WMAP team did, we consider a sharp cut-off at $k_c$, and set $P_{\Phi}(k')=0$ for $k'<k_c$.
However, $P_{\Phi}(k')$ of $k'<k_c$ may take on some non-zero value, though they are assumed to differ significantly from Eq. \ref{R1} and \ref{R2}. Therefore, our estimate on a transition scale should be interpreted as a lower bound, since a true $P_{\Phi,\mathrm{NL}}(k)$ is more likely to be higher than that of our sharp cut-off model, and a higher $k_c$ is needed to make 
a true $P_{\Phi,\mathrm{NL}}(k)$ equal to that of our sharp cut-off model. 
In our sharp cut-off model, Eq. \ref{P_NL2} is given by
\begin{eqnarray}
\lefteqn{P_{\Phi,\mathrm{NL}}(k)=\frac{(f_{\mathrm{NL}})^2}{\pi^2}A^2_0\int^{k_{max}}_{k_c}\frac{dk'}{k_0}\left(\frac{k+k'}{k_0}\right)^{n-4}}\label{P_NL2c}\\ 
&\times&[1+\epsilon_{\mathrm{TP}}\cos\theta(k+k')] 
\left(\frac{k'}{k_0}\right)^{n-2} [1+\epsilon_{\mathrm{TP}}\cos\theta(k')].\nonumber
\end{eqnarray}
We have found that $P_{\Phi,\mathrm{NL}}(k)$ of $k\gtrsim 2/\eta_0$ is barely affected by the value of $k_{\mathrm{max}}$, as long as $\log(k_{\mathrm{max}}/k_0)\gtrsim 5$. Hence, we have fixed $k_{\mathrm{max}}$ to $\log(k_{\mathrm{max}}/k_0)=10$, and numerically computed Eq. \ref{P_NL2c} by Romberg integration method \citep{Numerical_Recipe_C}.

\begin{figure}[htb!]
\centering\includegraphics[scale=.4]{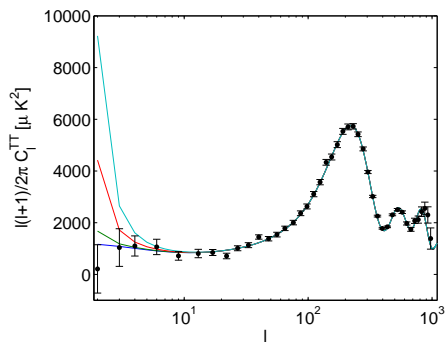}
\caption{CMB temperature power spectra of $\log(k_{\mathrm{c}}/k_0)=-120,-100,-60,-5$ (from the highest curve to the lowest), dots denote the WMAP and the ACBAR data.}
\label{Cl_TT}
\end{figure}
\begin{figure}[htb!]
\centering\includegraphics[scale=.4]{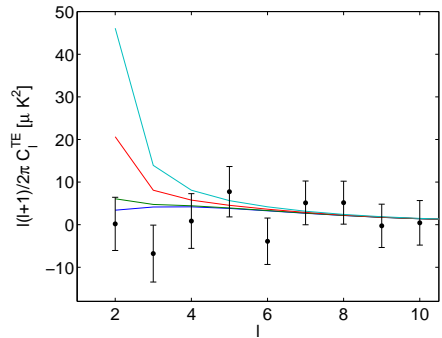}
\caption{CMB TE correlation of $\log(k_{\mathrm{c}}/k_0)=-120,-100,-60,-5$ (from the highest curve to the lowest),
dots denote the WMAP data.}
\label{Cl_TE}
\end{figure}
\begin{figure}[htb!]
\centering\includegraphics[scale=.4]{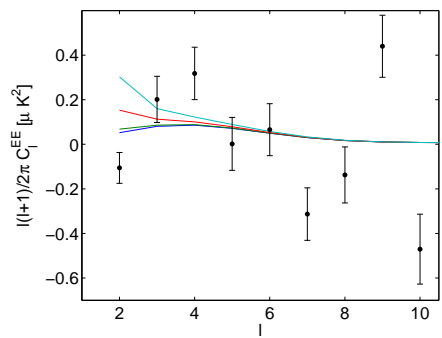}
\caption{E mode power spectrum of $\log(k_{\mathrm{c}}/k_0)=-120,-100,-60,-5$ (from the highest curve to the lowest),
 dots denote the WMAP data.}
\label{Cl_EE}
\end{figure}
\begin{figure}[htb!]
\centering\includegraphics[scale=.4]{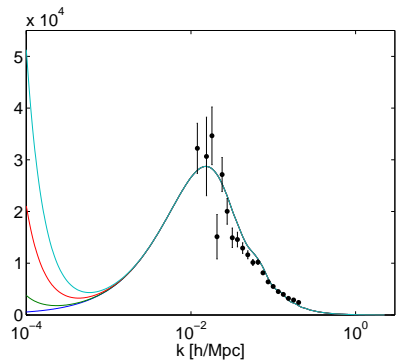}
\caption{matter power spectrum of $\log(k_{\mathrm{c}}/k_0)=-120,-100,-60,-5$ (from the highest to the lowest), dots denote SDSS data.}
\label{mpk}
\end{figure}

By making a small modification to \texttt{CAMB}, we have computed theoretical CMB and matter power spectrum, in which 
$P_{\Phi,\mathrm{NL}}(k)$ is taken into account. We show the theoretical CMB power spectra, TE correlation in Fig. \ref{Cl_TT}, \ref{Cl_TE} and \ref{Cl_EE}. The dots in the same plots denote the WMAP \citep{WMAP5:powerspectra} and the ACBAR data \citep{ACBAR2008}. We may see that anisotropy on largest scales ($l\lesssim 10$) is affected by $P_{\Phi,\mathrm{NL}}$ most.
For a E mode power spectrum and TE correlations, we show only low multipoles, since there is no visible effect on higher multipoles.
We show a theoretical matter power spectrum and SDSS data in Fig. \ref{mpk}. It also shows that matter inhomogeneities on largest scales ($k\lesssim 10^{-3} h/\mathrm{Mpc}$) are affected by $P_{\mathrm{NL}}$ most. As also noted by \citep{fnl_bias}, these excess power is, however, negligible on observable scales of the SDSS survey.
\begin{figure}[htb!]
\centering\includegraphics[scale=.35]{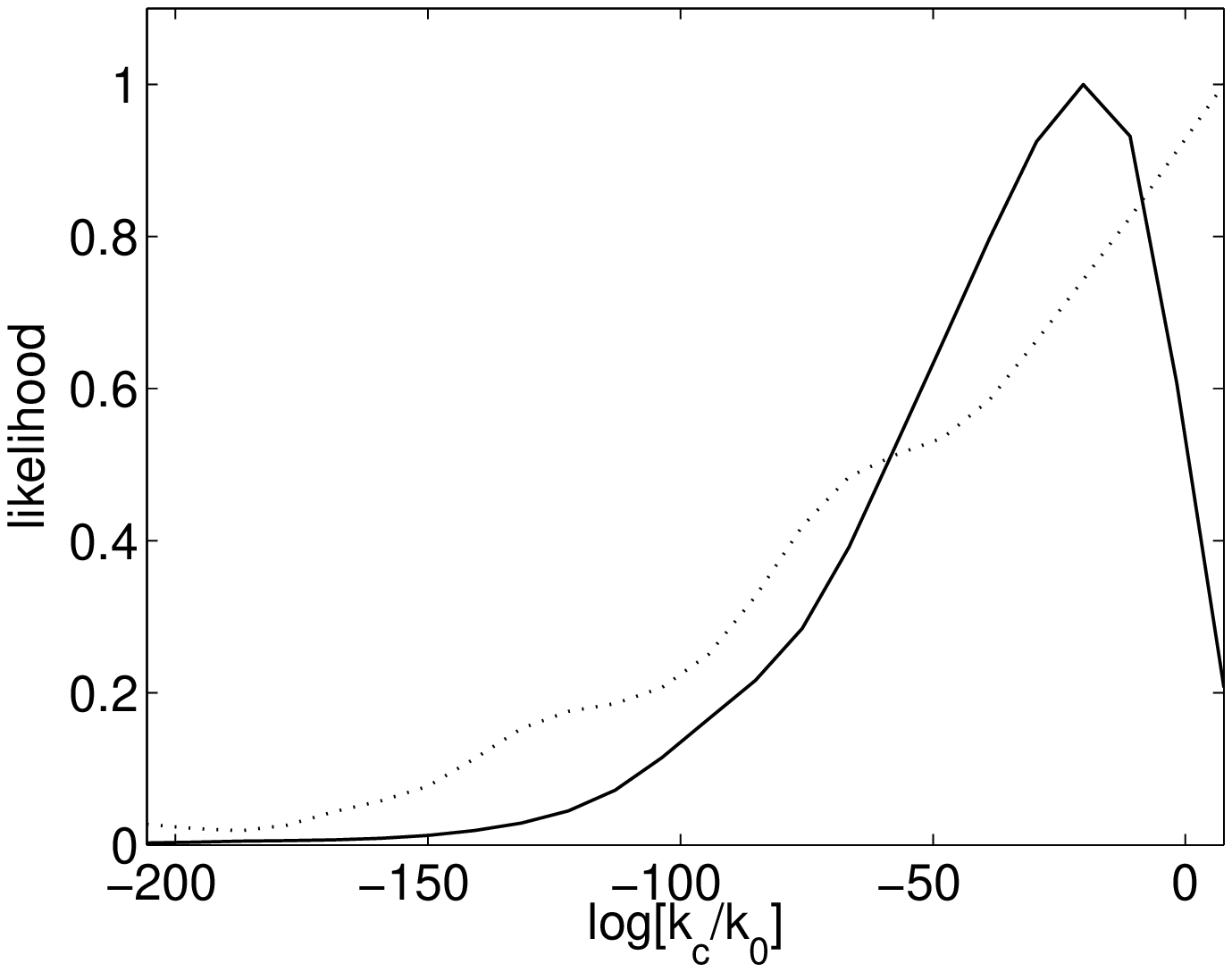}
\centering\includegraphics[scale=.35]{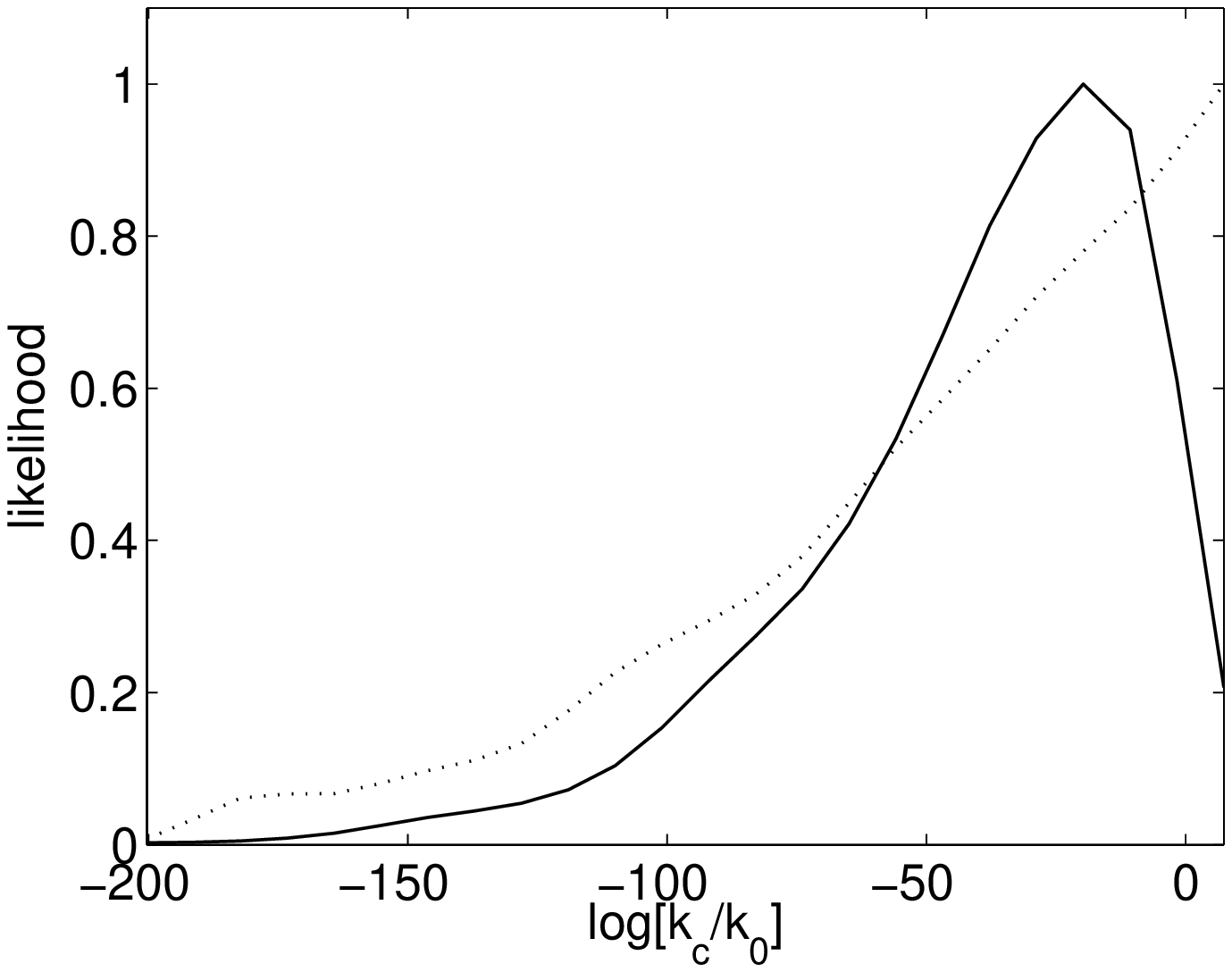}
\caption{Marginalized likelihood (solid lines) and mean likelihood of $\log(k_c/k_0)$ for the model I (top) and II (bottom)}
\label{L_kc}
\end{figure}

\begin{figure}[htb!]
\centering\includegraphics[scale=.6]{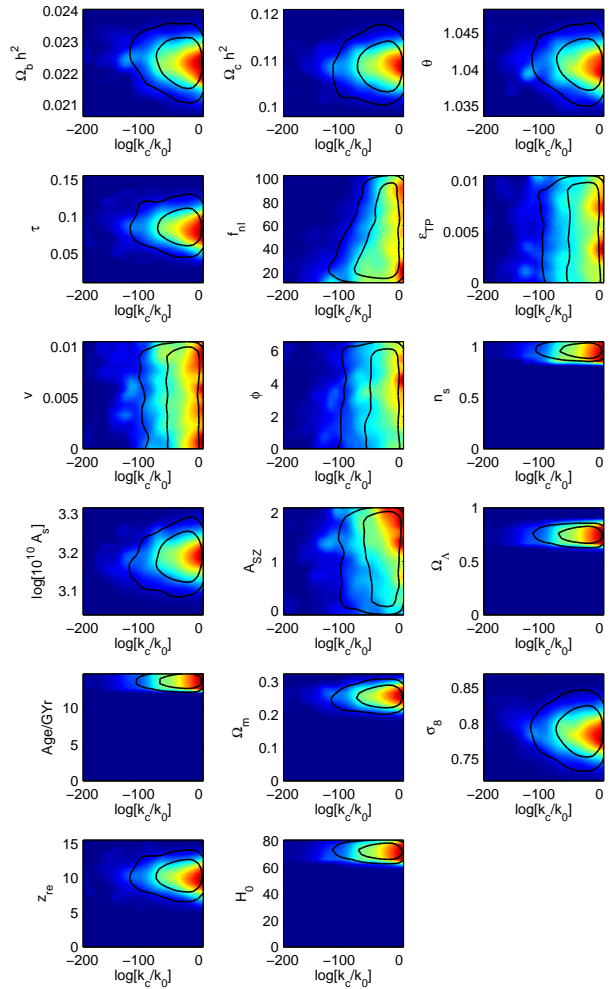}
\caption{Marginalized likelihood in the plane of $\log(k_c/k_0)$ versus others parameters for the model I. Two contour lines correspond to 1$\sigma$ and 2$\sigma$ levels.}
\label{L_2D1}
\end{figure}

\begin{figure}[htb!]
\centering\includegraphics[scale=.6]{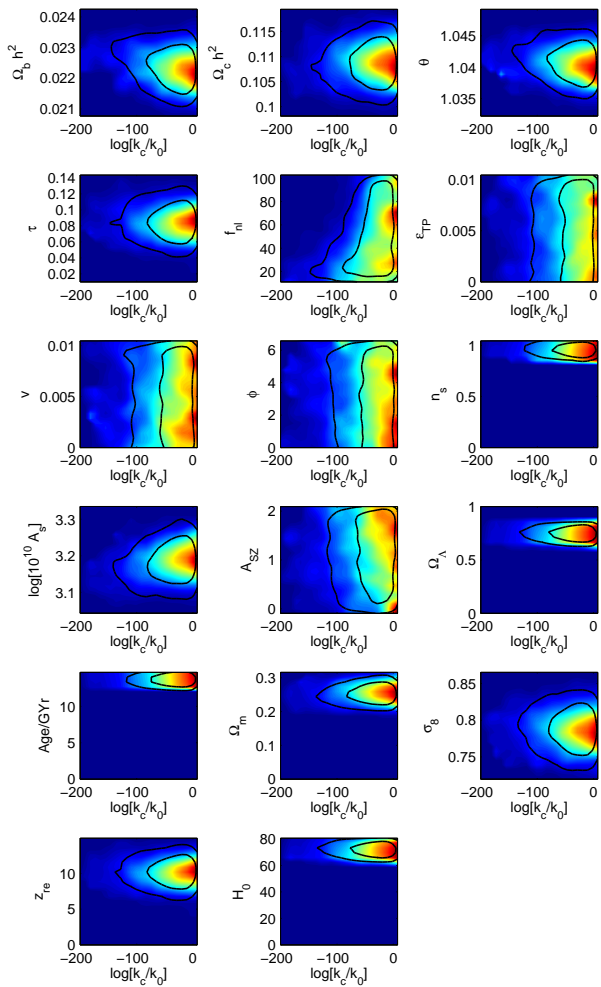}
\caption{Marginalized likelihood in the plane of $\log(k_c/k_0)$ versus others parameters for the model II. Two contour lines correspond to 1$\sigma$ and 2$\sigma$ levels.}
\label{L_2D2}
\end{figure}

Using a modified \texttt{CAMB} and \texttt{CosmoMC} \citep{CAMB,CosmoMC}, we have estimated $k_c$ respectively for the model I and II. For data constraints, we have used the SDSS data \cite{SDSS:tech,SDSS:low_galactic,SDSS:fifth_data}, the recent CMB observations (WMAP + ACBAR + QUaD \cite{WMAP5:basic_result,WMAP5:powerspectra,ACBAR,ACBAR2008,QUaD1,QUaD2,QUaD:instrument}), Supernovae data \citep{HST,SN_ESSENCE,SNLS} and Big-Bang Nucleosynthesis \citep{BBN}.
We show the marginalized likelihood (solid lines) and mean likelihood (dotted lines) distribution of $\log(k_c/k_0)$ in Fig. \ref{L_kc}. 
In Fig. \ref{L_2D1} and \ref{L_2D2}, we show the marginalized likelihood distribution in the plane of $\log(k_c/k_0)$ versus other parameters. The $k_c$ value of the best-fit cosmological model is $\log(k_c/k_0)=-1.98^{+0.35}_{-168.09}$ and $\log(k_c/k_0)=-1.98^{+0.35}_{-181.81}$ for the model I and II respectively. Note that the confidence interval is marginalized over $\epsilon_{\mathrm{TP}}$, $v$, $\phi$ and $f_{\mathrm{NL}}$ besides the basic $\Lambda$CDM parameters.
The best-fit values above do not coincide with the peak of likelihood distribution shown in Fig. \ref{L_kc}.
We attribute the discrepancy to the deviation of the multi-parameter likelihood function from Gaussian distribution.
The central values of our estimated $k_c$ are very similar to the cutoff scale found by the WMAP team \citep{WMAP3:parameter}.
Since CMB power spectra are sensitive to $P_{\Phi}(k)$ of $k\gtrsim 2/\eta_0$, $k_c$ higher than $2/\eta_0$ affects CMB power spectra through $P_{\Phi}(k)$ as well as $P_{\Phi,\mathrm{NL}}(k)$. This explains the similarity of the central values to the cutoff scale found by the WMAP team, even though $P_{\Phi,\mathrm{NL}}$ was not taken into account in their analysis. Note that the lower bounds above are associated mainly with $P_{\Phi,\mathrm{NL}}$, since $k_c\ll 2/\eta_0$. 

As shown in Fig. \ref{L_2D1} and \ref{L_2D2}, we find there is little degeneracy between $\log(k_c/k_0)$ and other parameters except for $A_{sz}$. As expected, the best-fit values of the basic $\Lambda$CDM parameters are similar to those of the WMAP concordance model. 

\subsection{scale-dependent $f_{\mathrm{NL}}$}
The nonlinear coupling parameter `$f_{\mathrm{NL}}$' is a local parameter, and hence possesses some scale-dependence \citep{fnl_review}.
In a single-field inflation, for instance,
$f_{\mathrm{NL}}$ in Eq. \ref{Phi_NL} is given by \citep{fnl_review}:
\begin{eqnarray}
f_{\mathrm{NL}} =\frac{5}{6}-3\frac{(\mathbf {k\cdot p})^2}{k^4}-\frac{2\mathbf {k\cdot p}-p^2 }{k^2}.\label{fnl_k}
\end{eqnarray}
In the models of a constant spectral index $n\sim 0.962$ and no transition, all terms of $f_{\mathrm{NL}}$ should have $k$ dependence $k^{\alpha\gtrsim 0.04}$ not to have unphysically large $P_{\mathrm{NL}}$.
However, $f_{\mathrm{NL}}$ predicted by most of inflationary models does not have such $k$ dependence.
Therefore, we find a scale-dependent $f_{\mathrm{NL}}$ alone does not provide a way to avoid unphysically large $P_{\mathrm{NL}}$.

\section{discussion}
\label{Discussion}
We have shown a primordial non-linear term (`$f_{\mathrm{NL}}$' term) may produce unphysically large CMB anisotropy, because of coupling to primordial fluctuation on largest scales.
Since such large excess power are not observed in CMB data, we have explored the following minimally extended power law models for a primordial power spectrum to explain the absence of the large excess power.
\begin{itemize}
\item 
A spectral index of a negative running: 
provided a power law model is valid up to the largest scale (i.e. no transition at a very low wavenumber), running of the spectral index should be negative (i.e. ${d\,n}/{d\ln k}<0$). We may rule out inflationary models of ${d\,n}/{d\ln k}\ge 0$ (e.g. a mass term potential and some models of softly broken SUSY models), 
\item
A transition at a very low wavenumber (e.g. cutoff): provided a spectral index is constant, there should exist some transition at a very low wavenumber, below which the power law is not valid. We have fitted a transition scale of a sharp cut-off model with the recent CMB and SDSS data, and obtained $\log(k_c/k_0)=-1.98^{+0.35}_{-168.09}$ and $\log(k_c/k_0)=-1.98^{+0.35}_{-181.81}$  respectively for two models described by Eq. \ref{R1} and \ref{R2}. 
\end{itemize}
Though it is not clear which condition is true for the primordial power spectrum, it is certain that at least one of two conditions should be met to avoid unphysically large CMB anisotropy. 

We shall be able to impose stronger constraints on inflationary models with the data from the upcoming PLANCK surveyor \citep{Planck:sensitivity,Planck:mission}.
The improved constraints on a running spectral index of scalar perturbation ${d\,n}/{d\ln k}$, tensor-to-scalar ratio $r$, and 
the spectral index of tensor perturbation $n_t$ will improves our understanding on inflation, and improves our understanding on how  
unphysically large $P_{\mathrm{NL}}$ is avoided.

\section{ACKNOWLEDGMENTS}
We are grateful to V. A. Rubakov for helpful discussion.
We acknowledge the use of the Legacy Archive for Microwave Background Data Analysis (LAMBDA), ACBAR, QUaD and SDSS data.
This work made use of the \texttt{CosmoMC} package.
This work was supported by FNU grant 272-06-0417, 272-07-0528 and 21-04-0355.

\bibliographystyle{unsrt}
\bibliography{/home/tac/jkim/Documents/bibliography}

\end{document}